\documentclass[english]{article}
\usepackage[T1]{fontenc}
\usepackage[latin1]{inputenc}
\usepackage{geometry}
\usepackage{amsmath}
\usepackage{setspace}
\usepackage{amssymb}

\makeatletter
\newcommand{\lyxaddress}[1]{
\par {\raggedright #1
\vspace{1.4em}
\noindent\par}
}

\usepackage{babel}
\makeatother
\begin{document}

\title{REVISITING QUATERNIONIC DUAL ELECTRODYNAMICS}

\author{P. S. Bisht$^{\text{(1)}}$ and O. P. S. Negi$^{\text{(1,2)}}$}

\maketitle

\lyxaddress{\begin{center}
$^{\text{(1)}}$ Department of Physics\\
Kumaun University\\
S. S. J. Campus\\
Almora - 263601 (India)
\par\end{center}}

\lyxaddress{\begin{center}
$^{\text{(2)}}$ Institute of Theoretical Physics\\
 Chinese Academy of Sciences\\
 KITP Building Room No.- 6304\\
Hai Dian Qu Zhong Guan Chun Dong Lu \\
55 Hao , Beijing - 100080, P.R.China
\par\end{center}}

\lyxaddress{\begin{center}
Email:- ps\_bisht123@rediffmail.com\\
ops\_negi@yahoo.co.in
\par\end{center}}

\begin{abstract}
Dual electrodynamics and corresponding Maxwell's equations ( in the
presence of monopole only) are revisited from dual symmetry and accordingly
the quaternionic reformulation of field equations and equation of
motion is developed in simple, compact and consistent manner. 
\end{abstract}

\section{Introduction}

Quaternions were invented by Hamilton \cite{key-1} in 1843 and Tait
\cite{key-2} promoted them in order to solve the problems of mathematics
and physics. Quaternions represent the natural extension of complex
numbers and form an algebra under addition and multiplication . The
reasons why Hamilton's quaternions have never become a prevalent formalism
in physics, while Hamilton's formulation of complex numbers has been
universally adopted, is an open question. The fact is that 'quaternion
structures' are very frequent in numerous areas of physics, the most
prominent examples being special relativity (i.e., Lorentz transformations),
electrodynamics, and spin. For this reason quaternions and their generalizations
keep reappearing in a number of forms, which are as numerous as diverse:
spinors, Einstein's semi vectors, Pauli matrices, Stokes parameters,
Eddington numbers, Clifford numbers, qubits, etc. On the other hand,
because quaternion algebra yields more efficient algorithms than matrix
algebra for three and four dimensional applications, their use in
computer simulations and graphics, numerical and symbolic calculations,
robotics, navigation, etc., has become more and more frequent in the
past few decades. A comprehensive list of references has already been
published in quaternion bibliographies \cite{key-3} on the practical
applications of quaternion analysis on various applied topics in theoretical
and mathematical physics.. The quaternionic formulation of electrodynamics
has along history \ref{eq:4}\ref{eq:5}\ref{eq:6}, stretching back
to Maxwell himself \cite{key-4} who used real (Hamilton) quaternion
in his original manuscript 'on the application of quaternion to electromagnetism'
and in his celebrated book {}``Treatise on Electricity and Magnetism''.
However, Maxwell \cite{key-4} used quaternion as the substitute of
common vector calculus which made his field equations difficult for
his contemporaries \cite{key-7} because the quaternionic formulation
in three - space brings several complications as the field of applicability
of real quaternions is Euclidean four - space \ref{eq:8} . Therefore,
the turning point, in using quaternions in theoretical physics, was
the creation of special relativity which unites space and time forming
a four - dimensional space - time.The formulation of physical laws
using real quaternions has then be replaced by complex ones, and it
has been recognized that complex quaternions represent a powerful
instrument in formulating classical physical laws. Complex (bi-) quaternions
form a division ring having a number of desirable properties that
allow the powerful theorems of modern algebra to be applied.Quaternion
analysis has since been rediscovered at regular intervals and accordingly
the Maxwell's Equations of electromagnetism were rewritten as one
quaternion equations \cite{key-9,key-10}. We have also \cite{key-11}
studied the quaternionic formulation for generalized electromagnetic
fields of dyons (particles carrying simultaneous existence of electric
and magnetic charges) in unique, simpler and compact notations.

Applying the electromagnetic duality in Maxwell's equation, in this
paper, we have discussed the dual electro dynamics, dual Maxwell's
equation, and equation of motion for dual electric charge (i.e. for
magnetic monopole). We have also reformulated the dynamical equations
of dual electrodynamics in terms of quaternion variables and accordingly
developed the dual quaternion electrodynamics in simple, compact and
consistent manner. It has been concluded that quaternionic formulation,
where the field equations reduce to a single quaternion equation,
has closed analogy between the sub and superluminal objects as the
norm of a quaternion four-vector behaves in the same manner as it
does under the influence of superluminal Lorentz transformations.

\section{Dual Electrodynamics}

The concept of electromagnetic (EM) duality has been receiving much
attention \cite{key-12} in gauge theories, field theories, supersymmetry
and super strings. Duality invariance is an old idea introduced a
century ago in classical electromagnetism for the following Maxwell's
equations in vacuum (using natural units $c=\hslash=1$, space-time
four-vector$\left\{ x^{\mu}\right\} =(t,x,y,z),$$\left\{ x_{\mu}\right\} =\eta_{\mu\nu}x^{\mu}$
and $\{\eta_{\mu\nu}=+1,-1,-1,-1=\eta^{\mu\nu}\}$ through out the
text),

\begin{eqnarray}
\overrightarrow{\nabla}\cdot\overrightarrow{E} & = & 0;\!\!\,\,\,\,\,\,\,\,\,\,\,\,\,\,\overrightarrow{\nabla}\cdot\overrightarrow{H}=0;\nonumber \\
\overrightarrow{\nabla}\times\overrightarrow{E} & =- & \frac{\partial\overrightarrow{H}}{\partial t};\!\!\,\,\,\,\,\,\,\,\,\,\,\,\,\,\overrightarrow{\nabla}\times\overrightarrow{H}=\frac{\partial\overrightarrow{E}}{\partial t};\label{eq:1}\end{eqnarray}
as these were invariant not only under Lorentz and conformal transformations
but also invariant under the following duality transformations,

\begin{eqnarray}
\overrightarrow{E} & \Rightarrow & \overrightarrow{E}\,\cos\vartheta+\overrightarrow{H}\,\sin\vartheta;\nonumber \\
\overrightarrow{H} & \Rightarrow & -\overrightarrow{E}\,\sin\vartheta+\overrightarrow{H}\,\cos\vartheta;\label{eq:2}\end{eqnarray}
where $\overrightarrow{E}$ and and $\overrightarrow{H}$ are respectively
the the electric and magnetic fields. For a particular value of $\vartheta=\frac{\pi}{2}$,
equations (\ref{eq:2}) reduces to

\begin{eqnarray}
\overrightarrow{E} & \rightarrow\overrightarrow{H};\,\,\,\,\,\,\, & \overrightarrow{H}\rightarrow-\overrightarrow{E};\label{eq:3}\end{eqnarray}
which can be written as 

\begin{eqnarray}
\left(\begin{array}{c}
\overrightarrow{E}\\
\overrightarrow{H}\end{array}\right) & \Rightarrow & \left(\begin{array}{cc}
0 & 1\\
-1 & 0\end{array}\right)\left(\begin{array}{c}
\overrightarrow{E}\\
\overrightarrow{H}\end{array}\right).\label{eq:4}\end{eqnarray}
Let us introduce a complex vector $\overrightarrow{\Psi}=\overrightarrow{E}+i\,\overrightarrow{H}$~$(i=\sqrt{-1})$
so that the Maxwell's equations (\ref{eq:1}) be written as 

\begin{eqnarray}
\overrightarrow{\nabla}\cdot\overrightarrow{\Psi} & = & 0;\!\!\,\,\,\,\,\,\,\,\,\,\,\,\,\,\overrightarrow{\nabla}\times\overrightarrow{\Psi}=i\frac{\partial\overrightarrow{\Psi}}{\partial t};\label{eq:5}\end{eqnarray}
which is also invariant under following duality transformations 

\begin{eqnarray}
\overrightarrow{\Psi} & \rightarrow & \exp(i\vartheta)\overrightarrow{\Psi}.\label{eq:6}\end{eqnarray}
The duality symmetry is lost if electric charge and current source
densities enter to the conventional Maxwell's equations given by 

\begin{eqnarray}
\overrightarrow{\nabla}\cdot\overrightarrow{E}=\rho & ; & \overrightarrow{\nabla}\times\overrightarrow{H}=\overrightarrow{j}+\frac{\partial\overrightarrow{E}}{\partial t}\Longrightarrow\partial_{\nu}F^{\mu\nu}=j^{\mu};\nonumber \\
\overrightarrow{\nabla}\cdot\overrightarrow{H}=0 & ; & \nabla\times\overrightarrow{E}=-\frac{\partial\overrightarrow{H}}{\partial t}\Longrightarrow\partial_{\nu}\widetilde{F^{\mu\nu}}=0;\label{eq:7}\end{eqnarray}
where $\left\{ j^{\mu}\right\} =(\rho,\,\overrightarrow{j})$ is described
as four-current source density. Consequently, Maxwell's equations
may be solved by introducing the concept of vector potential in either
two ways \cite{key-13}. The conventional choice has been used as 

\begin{eqnarray}
\overrightarrow{E}=-\frac{\partial\overrightarrow{A}}{\partial t}-grad\,\phi; & \,\,\,\,\,\overrightarrow{H}= & \overrightarrow{\nabla}\times\overrightarrow{A};\label{eq:8}\end{eqnarray}
where $\left\{ A^{\mu}\right\} =(\phi,\,\overrightarrow{A})$ is called
the four potential and accordingly the second pair (i. e. . $\overrightarrow{\nabla}\cdot\overrightarrow{H}=0;\nabla\times\overrightarrow{E}=-\frac{\partial\overrightarrow{H}}{\partial t}\Longrightarrow\partial_{\nu}\widetilde{F^{\mu\nu}}=0$)
of the Maxwell's equations (\ref{eq:7}) become kinematical identities
and the dynamics is contained in the first pair (i.e $\overrightarrow{\nabla}\cdot\overrightarrow{E}=\rho;\nabla\times\overrightarrow{H}=\overrightarrow{j}+\frac{\partial\overrightarrow{E}}{\partial t}\Longrightarrow\partial_{\nu}F^{\mu\nu}=j^{\mu}$)
and $\square A^{\mu}=j^{\mu}$ with the D'Alembertian $\square=\frac{\partial^{2}}{\partial t^{2}}-\frac{\partial^{2}}{\partial x^{2}}-\frac{\partial^{2}}{\partial y^{2}}-\frac{\partial^{2}}{\partial z^{2}}$.
Here, we have used the definition $F^{\mu\nu}=\partial^{\nu}A^{\mu}-\partial^{\mu}A^{\nu}$
for anti-symmetric electromagnetic field tensor whose components are
$F^{0j}=E_{j};$$F^{jk}=\varepsilon^{jkl}H_{l}\,(\forall j,k,l=1,2,3);\,\varepsilon^{jkl}=+1$
for cyclic , $\varepsilon^{jkl}=-1$ for anti-cyclic permutations,
$\varepsilon^{jkl}=0$ for repeated indices; $\widetilde{F^{\mu\nu}}=\frac{1}{2}\epsilon^{\mu\nu\lambda\omega}F_{\lambda\omega}(\forall\mu,\nu,\eta,\lambda=0,1,2,3);$
being the dual electromagnetic field tensor with $\widetilde{F^{0j}}=H_{j}$
and $\widetilde{F^{jk}}=\varepsilon^{jkl}E_{l}$; $\epsilon^{\mu\nu\lambda\omega}$as
the four dimensional generalization of $\varepsilon^{jkl}$. Equation
(\ref{eq:5}) is now modified as 

\begin{eqnarray}
\overrightarrow{\nabla}\cdot\overrightarrow{\Psi} & = & \rho;\qquad\overrightarrow{\nabla}\times\overrightarrow{\Psi}=i\frac{\partial\overrightarrow{\Psi}}{\partial t}+i\,\overrightarrow{j};\label{eq:9}\end{eqnarray}
which is no more invariant under duality transformations (\ref{eq:6}).
Here if we consider the another alternative way to write 

\begin{eqnarray}
\overrightarrow{H}=-\frac{\partial\overrightarrow{B}}{\partial t}-grad\,\varphi; & \qquad\overrightarrow{E}= & -\overrightarrow{\nabla}\times\overrightarrow{B};\label{eq:10}\end{eqnarray}
by introducing another potential $\left\{ B^{\mu}\right\} =(\varphi,\,\overrightarrow{B})$,
we see that source free Maxwell's equations (\ref{eq:1}) retain their
forms but Maxwell's equations (\ref{eq:7}) reduce to\begin{eqnarray}
\overrightarrow{\nabla}\cdot\overrightarrow{E}=0 & ;\,\,\,\,\,\, & \overrightarrow{\nabla}\times\overrightarrow{H}=\frac{\partial\overrightarrow{E}}{\partial t};\nonumber \\
\overrightarrow{\nabla}\cdot\overrightarrow{H}=\sigma & ;\,\,\,\, & \nabla\times\overrightarrow{E}=-\overrightarrow{\kappa}-\frac{\partial\overrightarrow{H}}{\partial t};\label{eq:11}\end{eqnarray}
where the first pair ($\overrightarrow{\nabla}\cdot\overrightarrow{E}=0;\overrightarrow{\nabla}\times\overrightarrow{H}=\frac{\partial\overrightarrow{E}}{\partial t}$)
becomes kinematical while the dynamics is contained in the second
pair ($\overrightarrow{\nabla}\cdot\overrightarrow{H}=\sigma;\nabla\times\overrightarrow{E}=-\overrightarrow{\kappa}-\frac{\partial\overrightarrow{H}}{\partial t}$).
Equation (\ref{eq:11}) may then be written in following covariant
forms

\begin{eqnarray}
\partial_{\nu}F^{\mu\nu}=0 & or & F_{\mu\nu,\nu}=0;\nonumber \\
\partial_{\nu}\widetilde{F^{\mu\nu}}=k^{\mu} & or & \widetilde{F_{\mu\nu,\nu}}=k_{\mu};\label{eq:12}\end{eqnarray}
where $\widetilde{F^{\mu\nu}}=\partial^{\nu}B^{\mu}-\partial^{\mu}B^{\nu}$;
$\widetilde{\widetilde{F^{\mu\nu}}}=F^{\mu\nu}$; $\left\{ k^{\mu}\right\} =(\sigma,\overrightarrow{\kappa})$
and $\left\{ k_{\mu}\right\} =(\sigma,-\overrightarrow{\kappa})$.
Equation (\ref{eq:11}) may also be obtained if we apply the transformations
(\ref{eq:3}) and (\ref{eq:4}) along with the following duality transformations
for potential and current i.e.

\begin{eqnarray}
A^{\mu} & \rightarrow B^{\mu}; & B^{\mu}\rightarrow-A^{\mu}\Longleftrightarrow\left(\begin{array}{c}
A^{\mu}\\
B^{\mu}\end{array}\right)\Rightarrow\left(\begin{array}{cc}
0 & 1\\
-1 & 0\end{array}\right)\left(\begin{array}{c}
A^{\mu}\\
B^{\mu}\end{array}\right);\nonumber \\
j^{\mu} & \rightarrow k^{\mu}; & k^{\mu}\rightarrow-j^{\mu}\Longleftrightarrow\left(\begin{array}{c}
j^{\mu}\\
k^{\mu}\end{array}\right)\Rightarrow\left(\begin{array}{cc}
0 & 1\\
-1 & 0\end{array}\right)\left(\begin{array}{c}
j^{\mu}\\
k^{\mu}\end{array}\right).\label{eq:13}\end{eqnarray}
So, we may identify the potential $\left\{ B^{\mu}\right\} =(\varphi,\,\overrightarrow{B})$
as the dual potential and the current $\left\{ k^{\mu}\right\} =(\sigma,\overrightarrow{\kappa})$
as the dual current. Correspondingly the differential equations (\ref{eq:11})
are identified as the dual Maxwell's equations and we may accordingly
develop the dual electrodynamics. Introducing the electromagnetic
duality, in the Maxwell's equations , we may establish the connection
between electric and magnetic charge (monopole) \cite{key-17,key-18}
, with the fact that an electric charge interacts with its electric
field as the dual charge (magnetic monopole) interacts with magnetic
field, as,

\begin{eqnarray}
e & \rightarrow & g;\,\,\, g\rightarrow-e\Longleftrightarrow\left(\begin{array}{c}
e\\
g\end{array}\right)\Rightarrow\left(\begin{array}{cc}
0 & 1\\
-1 & 0\end{array}\right)\left(\begin{array}{c}
e\\
g\end{array}\right)\label{eq:14}\end{eqnarray}
where $g$ is described as the dual charge (charge of magnetic monopole).
Hence we may recall the dual electrodynamics as the dynamics of magnetic
monopole and the corresponding physical variables associated there
are described as the dynamical quantities of magnetic monopole. We
may also write the duality transformation for $F^{\mu\nu}$ and $\widetilde{F^{\mu\nu}}$
as

\begin{eqnarray}
F^{\mu\nu}\rightarrow\widetilde{F^{\mu\nu}}\,;\quad\widetilde{F^{\mu\nu}}\rightarrow F^{\mu\nu} & \Longleftrightarrow & \left(\begin{array}{c}
F^{\mu\nu}\\
\widetilde{F^{\mu\nu}}\end{array}\right)\Rightarrow\left(\begin{array}{cc}
0 & 1\\
-1 & 0\end{array}\right)\left(\begin{array}{c}
F^{\mu\nu}\\
\widetilde{F^{\mu\nu}}\end{array}\right).\label{eq:15}\end{eqnarray}
So, we may rewrite the dual Maxwell's equations (\ref{eq:11}) as
the field equations for magnetic monopole (or in the absence of electric
charge) on replacing $\sigma$ by magnetic charge density $\rho_{m}$
and $\overrightarrow{\kappa}$ by the magnetic current density $\overrightarrow{j_{m}}$
as 

\begin{eqnarray}
\overrightarrow{\nabla}.\overrightarrow{E} & = & 0;\nonumber \\
\overrightarrow{\nabla}.\overrightarrow{H} & = & \rho_{m};\nonumber \\
\overrightarrow{\nabla}\times\overrightarrow{E} & = & -\frac{\partial\overrightarrow{H}}{\partial t}-\overrightarrow{j_{m}};\nonumber \\
\overrightarrow{\nabla}\times\overrightarrow{H} & = & \frac{\partial\overrightarrow{E}}{\partial t}.\label{eq:16}\end{eqnarray}
The complex vector field $\overrightarrow{\Psi}=\overrightarrow{E}+i\,\overrightarrow{H}(i=\sqrt{-1})$
is now replaced by $\overrightarrow{\psi}$ as the consequence of
duality (\ref{eq:3}) and (\ref{eq:4}) i.e.

\begin{eqnarray}
\overrightarrow{\psi} & = & \overrightarrow{H}-i\overrightarrow{E}.\label{eq:17}\end{eqnarray}
Substituting the $\overrightarrow{E}$ and $\overrightarrow{H}$ from
equations (\ref{eq:10}) on replacing $\varphi$by $\phi_{m}$ we
may establish the following relation between the electromagnetic vector
$\overrightarrow{\psi}$ and the components of the magnetic four -
potential as,

\begin{eqnarray}
\overrightarrow{\psi} & = & -\overrightarrow{\nabla}\phi_{m}-\frac{\partial\overrightarrow{B}}{\partial t}+i\overrightarrow{\nabla}\times\overrightarrow{B}.\label{eq:18}\end{eqnarray}
Hence we may write the Maxwell's equations (\ref{eq:16}) for monopole
in terms of complex vector field $\overrightarrow{\psi}$ as

\begin{eqnarray}
\overrightarrow{\nabla}.\overrightarrow{\psi} & = & \rho_{m},\nonumber \\
\overrightarrow{\nabla}\times\overrightarrow{\psi} & = & i\overrightarrow{j_{m}}+i\frac{\partial\overrightarrow{\psi}}{\partial t}.\label{eq:19}\end{eqnarray}
Accordingly let us replace dual $\widetilde{F^{\mu\nu}}$ by the field
tensor $\mathcal{F}^{\mu\nu}$ for magnetic monopole as,

\begin{eqnarray}
\mathcal{F}_{\mu\nu} & = & \partial_{\nu}B_{\mu}-\partial_{\mu}B_{\nu}\,(\mu,\nu=1,2,3)\label{eq:20}\end{eqnarray}
which reproduces the following definition of magneto-electric fields
of monopole as

\begin{eqnarray}
\mathcal{F}_{0i} & = & H^{i},\nonumber \\
\mathcal{F}_{ij} & = & -\varepsilon_{ijk}E^{k}.\label{eq:21}\end{eqnarray}
Hence the covariant form of Maxwell's equations (\ref{eq:12}) for
magnetic monopole may now be written as

\begin{eqnarray}
\mathcal{F_{\mathit{\mu\nu,\nu}}} & = & k_{\mu},\nonumber \\
\widetilde{\mathcal{F}_{\mu\nu,\nu}} & = & 0\label{eq:22}\end{eqnarray}
where $\{ k_{\mu}\}=\{\rho_{m},-\overrightarrow{j_{m}}\}$ is the
four - current density of the magnetic charge. Now using equations
(\ref{eq:10}) and (\ref{eq:16}), we get 

\begin{eqnarray}
\square\phi_{m} & = & \rho_{m};\nonumber \\
\square\overrightarrow{B} & = & \overrightarrow{j_{m}};\label{eq:23}\end{eqnarray}
where we have imposed the following Lorentz gauge condition

\begin{eqnarray}
\overrightarrow{\nabla}.\overrightarrow{B}+\frac{\partial\phi_{m}}{\partial t} & = & 0.\label{eq:24}\end{eqnarray}
Equation (\ref{eq:23}) can also be generalized in the following covariant
form, which may directly be obtained from equation (\ref{eq:22}),
as

\begin{eqnarray}
\square B_{\mu} & = & k_{\mu}\label{eq:25}\end{eqnarray}
Consequently, the divergence of the third equation (\ref{eq:16})
( i.e. $div$ of $[\overrightarrow{\nabla}\times\overrightarrow{E}=-\frac{\partial\overrightarrow{H}}{\partial t}-\overrightarrow{j_{m}}]$)
leads to the following continuity equation of magnetic charge (dual
electrodynamics),

\begin{eqnarray}
\overrightarrow{\nabla}.\overrightarrow{j_{m}}+\frac{\partial\rho_{m}}{\partial t} & = & 0.\label{eq:26}\end{eqnarray}
 Taking the curl of , $\overrightarrow{\nabla}\times\overrightarrow{\psi}=i\overrightarrow{j_{m}}+i\frac{\partial\overrightarrow{\psi}}{\partial t}$
, the second set and using, $\overrightarrow{\nabla}.\overrightarrow{\psi}=\rho_{m}$,
the first set of equation (\ref{eq:19}), we get the following differential
equation 

\begin{eqnarray}
\square\overrightarrow{\psi} & = & -\overrightarrow{\nabla}\rho_{m}-\frac{\partial\overrightarrow{k}}{\partial t}+i\overrightarrow{\nabla}\times\overrightarrow{j_{m}}=\overrightarrow{S}\,\,(say)\label{eq:27}\end{eqnarray}
where $\overrightarrow{S}$ is being introduced as a new parameter
which we may call a field current. As such, we have established a
connection between the field vector $\overrightarrow{\psi}$ and field
current $\overrightarrow{S}$ in the same manner as we have established
the relation (\ref{eq:23} or \ref{eq:25}) between the potential
and current. Accordingly, we may develop the classical Lagrangian
formulation in order to obtain the field equation (dual Maxwell's
equations) and equation of motion for the dynamics of a dual charge
(magnetic monopole) inter acing with electromagnetic field.

The Lorentz force equation of motion for a dual charge (i.e magnetic
monopole) may now be written from the duality equations (\ref{eq:3})
and (\ref{eq:14}) as 

\begin{eqnarray}
\frac{d\overrightarrow{p}}{d\tau}=\overrightarrow{f}=m & \overrightarrow{\ddot{x}}= & g(\overrightarrow{H}\,-\,\overrightarrow{v}\times\overrightarrow{E})\label{eq:28}\end{eqnarray}
where $\overrightarrow{p}=m\,\overrightarrow{\dot{x}}=m\,\overrightarrow{v}$
is the momentum, and $\overrightarrow{f}$ is a force acting on a
particle of charge $g$, mass $m$ and moving with the velocity $\overrightarrow{v}$
in electromagnetic fields. Equation (\ref{eq:28}) can be generalized
to write it in the following four vector formulation as

\begin{eqnarray}
\frac{dp^{\mu}}{d\tau} & =f^{\mu}=m\ddot{x}_{\mu} & =g\mathcal{F^{\mu\nu}}U_{\nu}\label{eq:29}\end{eqnarray}
where $\left\{ U_{\nu}\right\} =\dot{x_{\nu}}$ is the four velocity
while $\ddot{x}_{\mu}$is the four-acceleration of a particle.

\section{Quaternion Dual Electrodynamics}

In order to obtain the field equations and equation of motion of a
dual charge (i.e a magnetic monopole), we may now revisit to the quaternionic
reformulation of the dual electrodynamics. A complex quaternion (
bi-quaternion) is expressed as

\begin{eqnarray}
\mathsf{Q} & = & Q_{\mu}e_{\mu}\,\,\,(Q_{\mu}\in\mathbb{C})\label{eq:30}\end{eqnarray}
where $Q_{\mu}$ are complex quantities for $(\mu=0,1,2,3)$ while
$e_{0}=1$, $e_{j}(\forall j=1,2,3)$ are the quaternion units satisfying
the following multiplication rules,

\begin{eqnarray}
e_{0}e_{j}=e_{j}e_{0}=e_{j};\qquad e_{j}e_{k} & = & -\delta_{jk}+\varepsilon_{jkl}e_{l};\,\,\,\,(\forall j,k,l=1,2,3)\label{eq:31}\end{eqnarray}
where $\delta_{jk}$ is Kronecker delta and $\varepsilon_{jkl}$ is
the three index Levi - Civita symbol. A real or complex quaternion
can be written as $\mathsf{Q}=Q_{0}+Q_{j}e_{j}\Rightarrow Q_{0}+\overrightarrow{Q}$
as the combination of a scalar a vector (i.e $\mathsf{Q}=(Q_{0},\overrightarrow{Q})$
with $Q_{0}e_{0}=Q_{0}$and $\overrightarrow{Q}=Q_{1}e_{1}+Q_{2}e_{2}+Q_{3}e_{3}$
). The sum and product of two quaternions $\mathsf{a}=(a_{0},\,\overrightarrow{a})$and
$\mathsf{b}=(b_{0},\overrightarrow{\, b})$ are respectively defined
as $(a_{0},\,\overrightarrow{a})+(b_{0},\overrightarrow{\, b})=(a_{0}+b_{0},\,\overrightarrow{a}+\overrightarrow{b})$
and $(a_{0},\,\overrightarrow{a})\cdot(b_{0},\overrightarrow{\, b})=(a_{0}b_{0}-\overrightarrow{a}\cdot\overrightarrow{b}\,,\,\overrightarrow{a}\times\overrightarrow{b}+a_{0}\overrightarrow{b}+b_{0}\overrightarrow{a})$.
The norm of a quaternion is $N(\mathsf{Q})=\mathsf{Q}\,\overline{\mathsf{Q}}=\overline{\mathsf{Q}}\,\mathsf{Q}=Q_{0}^{2}+Q_{1}^{2}+Q_{2}^{2}+Q_{3}^{2}$;
where $\overline{\mathsf{Q}}=(Q_{0}\,,\,-\overrightarrow{Q})$ (as
$\overline{e_{j}}=-e_{j};$$\overline{e_{0}}=e_{0}=1$ ) is the quaternion
conjugate which follows the law of anti- automorphism i.e. $\overline{(\mathsf{p}\mathsf{q})}=\overline{\mathsf{q}}\,\overline{\mathsf{p}}$
where $\mathsf{p}$ and $\mathsf{q}$ are two quaternions. Quaternions
are associative but anti-commutative in nature and form a group as
well as a division ring. Quaternions have the inverse $\mathsf{Q}^{-1}=\frac{\overline{\mathsf{Q}}}{N(\mathsf{Q})}$.
So, quaternions are equivalent to the four dimensional representation
of our space-time. We can divide a quaternion by other quaternion
resulting to a third quaternion. There is no distinction between the
contravariants and covariants if we write a four vector in quaternion
representation. A real quaternion may be expressed as a four vector
in Euclidean space with signature $(+,+,+,+)$ while the bi-quaternion
may be written in four-dimensional Minikowski space and the signature
is chosen as $(-,+,+,+)$. For contraction and expansion of ranks
we may use the products $\left\langle \mathsf{p}\,,\,\mathsf{q}\right\rangle _{Sc}=\frac{1}{2}(\mathsf{p}\,\overline{\mathsf{q}}+\mathsf{q}\,\mathsf{\overline{p}})$
(scalar product ) and $\left\langle \mathsf{p}\,,\,\mathsf{q}\right\rangle _{Vec}=\frac{1}{2}(\mathsf{p}\,\overline{\mathsf{q}}-\mathsf{q}\,\overline{\mathsf{p}})$
(vector product) of two quaternions. As such, we may write the quaternion
analogue of a space-time contravariant four vector as 

\begin{eqnarray}
x^{\mu}\mapsto\mathcal{\mathfrak{\mathbb{\mathit{\mathrm{\mathsf{x}}}}}}= & -it\,+\,\overrightarrow{x}\Leftrightarrow & -it+x_{1}e_{1}+x_{2}e_{2}+x_{3}e_{3}\label{eq:32}\end{eqnarray}
and the covariant four vector may be written in terms of analogous
quaternion representation as

\begin{eqnarray}
x_{\mu}\mapsto\bar{\mathbf{\mathrm{\mathsf{x}}}}= & -it\,-\,\overrightarrow{x}\Leftrightarrow & -it-x_{1}e_{1}-x_{2}e_{2}-x_{3}e_{3}.\label{eq:33}\end{eqnarray}
So, the quaternionic four dimensional Nabla (four differential operator)
and its conjugate representations are written as 

\begin{eqnarray}
\mathsf{\mathbf{\mathrm{\mathbf{\boldsymbol{\boxdot}}}}} & = & i\partial_{t}+e_{1}\partial_{1}+e_{2}\partial_{2}+e_{3}\partial_{3};\label{eq:34}\\
\overline{\mathbf{\mathsf{\boldsymbol{\mathbf{\boxdot}}}}} & = & i\partial_{t}-e_{1}\partial_{1}-e_{2}\partial_{2}-e_{3}\partial_{3};\label{eq:35}\end{eqnarray}
where $\partial_{t}=\frac{\partial}{\partial t},\,\,\partial_{1}=\frac{\partial}{\partial x_{1}}=\partial_{x},\,\,\partial_{2}=\frac{\partial}{\partial x_{2}}=\partial_{y},\,\,\partial_{3}=\frac{\partial}{\partial x_{3}}=\partial_{z}.$
Hence, straight forwardly, we may write the following quaternionic
forms respectively analogous to dual potential $\left\{ B^{\mu}\right\} =(\phi_{m},\,\overrightarrow{B})$
and dual current $\left\{ k^{\mu}\right\} =(\rho_{m},\overrightarrow{j_{m}})$
associated with monopole as, 

\begin{eqnarray}
\mathfrak{\mathcal{\mathrm{\mathrm{\mathsf{B}}}}} & = & i\phi_{m}+e_{1}B_{1}+e_{2}B_{2}+e_{3}B_{3};\label{eq:36}\\
\mathsf{k} & = & i\rho_{m}+e_{1}j_{m1}+e_{2}j_{m2}+e_{3}j_{m3}.\label{eq:37}\end{eqnarray}
Now operating quaternionic Nabla, given by equation (\ref{eq:34}),
respectively on equations (\ref{eq:36}) and (\ref{eq:37}) and using
the multiplication rules of quaternion units given by equation (\ref{eq:31}),
we get;

\begin{eqnarray}
\boldsymbol{\mathrm{\mathbf{\boxdot}}}\mathsf{B} & = & -(\partial_{t}\phi_{m}+\partial_{1}B_{1}+\partial_{2}B_{2}+\partial_{3}B_{3})\nonumber \\
-i & e_{1} & [-\partial_{t}B_{1}-\partial_{1}\phi_{m}+i(\partial_{2}B_{3}-\partial_{3}B_{2})]\nonumber \\
-i & e_{2} & [-\partial_{t}B_{2}-\partial_{2}\phi_{m}+i(\partial_{3}B_{1}-\partial_{1}B_{3})]\nonumber \\
-i & e_{3} & [-\partial_{t}B_{3}-\partial_{3}\phi_{m}+i(\partial_{1}B_{2}-\partial_{2}B_{1})]\label{eq:38}\end{eqnarray}
and

\begin{eqnarray}
\mathbf{\boldsymbol{\boxdot}}\mathsf{k} & = & -(\partial_{t}\rho_{m}+\partial_{1}j_{m1}+\partial_{2}j_{m2}+\partial_{3}j_{m3})\nonumber \\
-i & e_{1} & [-\partial_{t}j_{m1}+\partial_{1}\rho_{m}+i(\partial_{2}j_{m3}-\partial_{3}j_{m2})]\nonumber \\
-i & e_{2} & [-\partial_{t}j_{m2}+\partial_{2}\rho_{m}+i(\partial_{3}j_{m1}-\partial_{1}j_{m3})]\nonumber \\
-i & e_{3} & [-\partial_{t}j_{m3}+\partial_{3}\rho_{m}+i(\partial_{1}j_{m2}-\partial_{2}j_{m1})].\label{eq:39}\end{eqnarray}
Comparing these equations with equations (\ref{eq:18}) and (\ref{eq:27}),
we get

\begin{eqnarray}
\boldsymbol{\mathrm{\mathbf{\boxdot}}}\mathsf{B} & = & \psi_{0}-ie_{1}\psi_{1}-ie_{2}\psi_{2}-ie_{3}\psi_{3};\label{eq:40}\\
\mathbf{\boldsymbol{\boxdot}}\mathsf{k} & = & S_{0}-ie_{1}S_{1}-ie_{2}S_{2}-ie_{3}S_{3};\label{eq:41}\end{eqnarray}
where 

\begin{eqnarray}
\psi_{0} & = & -(\partial_{t}\phi_{m}+\partial_{1}B_{1}+\partial_{2}B_{2}+\partial_{3}B_{3})=0;\label{eq:42}\\
S_{0} & = & -(\partial_{t}\rho_{m}+\partial_{1}j_{m1}+\partial_{2}j_{m2}+\partial_{3}j_{m3})=0;\label{eq:43}\end{eqnarray}
after applying the Lorentz gauge condition (\ref{eq:24}) for $\psi_{0}=0$
and continuity equation (\ref{eq:26}) for $S_{0}=0$. So, we obtain
the following compact and simpler forms of quaternionic inhomogeneous
wave equations for potential and current for dual charge (magnetic
monopole),

\begin{eqnarray}
\mathbf{\boldsymbol{\boxdot}}\mathbf{B} & = & \mathrm{\mathit{\mathfrak{\mathsf{\mathfrak{\mathcal{\boldsymbol{\psi}};}}}}}\label{eq:44}\\
\mathbf{\boldsymbol{\boxdot}}\mathsf{k} & = & \mathsf{S};\label{eq:45}\end{eqnarray}
where $\mathbf{\boldsymbol{\psi}}$ and $\mathsf{S}$ are respectively
identified as the quaternionic forms of generalized field ( four-field)
and field density (four-field-current) given by 

\begin{eqnarray}
\mathbf{\mathbf{\boldsymbol{\psi}}} & = & \psi_{0}-ie_{1}\psi_{1}-ie_{2}\psi_{2}-ie_{3}\psi_{3};\label{eq:46}\\
\mathsf{S} & = & S_{0}-ie_{1}S_{1}-ie_{2}S_{2}-ie_{3}S_{3}.\label{eq:47}\end{eqnarray}
As such, the quaternion wave equations \eqref{eq:44} and \eqref{eq:45}
are regarded as the quaternion field equations for potential and currents
associated with monopoles. These quaternion field equations are invariant
under quaternion and Lorentz transformations. Similarly, if we operate
\eqref{eq:35} to \eqref{eq:46} , we get

\begin{eqnarray}
\overline{\mathbf{\mathsf{\mathbf{\boldsymbol{\boxdot}}}}}\,\boldsymbol{\psi} & = & (-i\overrightarrow{\nabla}\cdot\overrightarrow{\psi})-\sum_{j=1}^{3}e_{j}[(\partial_{t}\psi_{j}+i(\overrightarrow{\nabla}\times\overrightarrow{\psi})_{j}].\label{eq:48}\end{eqnarray}
which is reduced to the following quaternionic wave equation on using
equations \eqref{eq:19} and \eqref{eq:37} i.e. 

\begin{eqnarray}
\overline{\mathbf{\mathsf{\mathbf{\boldsymbol{\boxdot}}}}}\,\boldsymbol{\psi} & = & -i\rho_{m}-e_{1}j_{m1}-e_{2}j_{m2}-e_{3}j_{m3}=-\mathsf{k}\label{eq:49}\end{eqnarray}
showing the relation between fields and current and is thus analogous
to the Maxwell's field equations \eqref{eq:16} or \eqref{eq:19}
in quaternionic formulation. Now using quaternion field equations
\eqref{eq:44} and \eqref{eq:45}, we get \begin{eqnarray}
\overline{\mathbf{\mathsf{\mathbf{\boldsymbol{\boxdot}}}}}\,\boldsymbol{\psi}=\overline{\boldsymbol{\boxdot}}(\boldsymbol{\boxdot}\mathsf{B)} & =(\overline{\boldsymbol{\boxdot}}\boldsymbol{\boxdot})\,\mathsf{B=-\square}\mathsf{B=} & -\mathsf{k};\label{eq:50}\end{eqnarray}
\begin{eqnarray}
\mathbf{\boldsymbol{\boxdot}}\,\boldsymbol{\mathsf{k}}=\boldsymbol{\boxdot}(-\overline{\boldsymbol{\boxdot}} & \boldsymbol{\psi})=-(\boldsymbol{\boxdot}\mathsf{\overline{\boldsymbol{\boxdot}})\boldsymbol{\psi}=\square}\mathsf{\boldsymbol{\psi}=} & \mathsf{S};\label{eq:51}\end{eqnarray}
which may also be written as 

\begin{eqnarray}
\boldsymbol{\boxdot}\overline{\boldsymbol{\boxdot}} & \mathsf{B}= & -\square\mathsf{B=-\mathsf{k}};\label{eq:52}\\
\boldsymbol{\boxdot}\overline{\boldsymbol{\boxdot}} & \mathsf{\boldsymbol{\psi}}= & -\square\mathsf{\boldsymbol{\psi}=-\mathsf{S}.}\label{eq:53}\end{eqnarray}
Equations \eqref{eq:52} and \eqref{eq:53} are analogous to equations
\eqref{eq:25} and \eqref{eq:27} in simple,compact and consistent
quaternion formulation and the D'Alembertian operator $\square=\frac{\partial^{2}}{\partial t^{2}}-\frac{\partial^{2}}{\partial x^{2}}-\frac{\partial^{2}}{\partial y^{2}}-\frac{\partial^{2}}{\partial z^{2}}=\boldsymbol{-\boxdot}\overline{\boldsymbol{\boxdot}}=-\overline{\boldsymbol{\boxdot}}\boldsymbol{\boxdot}$
is described as the negative of modulus of quaternion Nabla while
if we define it in Minikowski space of signature $(-,+,+,+)$ it becomes
the norm of quaternion Nabla $N(\boldsymbol{\boxdot})=\overline{\boldsymbol{\boxdot}}\boldsymbol{\boxdot}=\boldsymbol{\boxdot}\overline{\boldsymbol{\boxdot}}=\boldsymbol{\square}$.
As such, we may express the quaternionic forms of the Lorentz force
$\left\{ f_{\mu}\right\} $, the velocity $\left\{ U_{\nu}\right\} $and
the field tensor $\left\{ \mathcal{F}_{\mu\nu}\right\} $as

\begin{eqnarray}
\boldsymbol{\mathbf{\Im}} & = & f_{0}+e_{1}f_{1}+e_{2}f_{2}+e_{3}f_{3}=(f_{0}\,,\overrightarrow{f\,});\label{eq:54}\\
\mathsf{U} & = & U_{0}+e_{1}U_{1}+e_{2}U_{2}+e_{3}U_{3}=(U_{0}\,,\overrightarrow{U\,});\label{eq:55}\\
\boldsymbol{\mathcal{\boldsymbol{F}}} & =\mathcal{F_{\mu\nu}}e_{\nu}=(\digamma_{0},\overrightarrow{\digamma})= & \boldsymbol{\digamma}_{\mu0}e_{0}+\boldsymbol{\digamma}_{\mu1}e_{1}+\boldsymbol{\digamma}_{\mu2}e_{2}+\boldsymbol{\digamma}_{\mu3}e_{3};\label{eq:56}\end{eqnarray}
where 

\begin{eqnarray}
\boldsymbol{\digamma_{0}=\digamma}_{\mu0} & = & \digamma_{00}e_{0}+\digamma_{10}e_{1}+\digamma_{20}e_{2}+\digamma_{30}e_{3};\nonumber \\
\boldsymbol{\digamma_{1}=\digamma}_{\mu1} & = & \digamma_{01}e_{0}+\digamma_{11}e_{1}+\digamma_{21}e_{2}+\digamma_{31}e_{3};\nonumber \\
\boldsymbol{\digamma}_{\mathbf{2}}=\boldsymbol{\digamma}_{\mu2} & = & \digamma_{02}e_{0}+\digamma_{12}e_{1}+\digamma_{22}e_{2}+\digamma_{32}e_{3};\nonumber \\
\boldsymbol{\digamma}_{\mathbf{3}}=\boldsymbol{\digamma}_{\mu3} & = & \digamma_{03}e_{0}+\digamma_{13}e_{1}+\digamma_{23}e_{2}+\digamma_{33}e_{3}.\label{eq:57}\end{eqnarray}
As such, we see from equations (\ref{eq:56}) and (\ref{eq:57}) that
the four components of second rank antisymmetric tensor $\mathcal{F_{\mu\nu}}$are
also quaternions. Hence, we may write the quaternion form of covariant
Maxwell's equations (\ref{eq:22}) in the following manner,

\begin{eqnarray}
[\boldsymbol{\boxdot},\,\boldsymbol{\mathcal{\boldsymbol{F}}}] & = & \boldsymbol{\mathsf{k}}\label{eq:58}\end{eqnarray}
where $[\boldsymbol{\boxdot},\,\boldsymbol{\mathcal{\boldsymbol{F}}}]$
is the scalar product derived in the following manner,

\begin{eqnarray}
[\boldsymbol{\boxdot},\,\boldsymbol{\mathcal{\boldsymbol{F}}}] & =\frac{1}{2} & (\overline{\boldsymbol{\boxdot}}\boldsymbol{\mathcal{\boldsymbol{F}}}+\overline{\boldsymbol{\mathcal{\boldsymbol{F}}}}\boldsymbol{\underleftarrow{\boxdot}})\nonumber \\
= & \partial_{0}\boldsymbol{\digamma_{0}}+ & \overrightarrow{\nabla}\cdot\boldsymbol{\overrightarrow{\digamma}}=(i\,\rho_{m},\,\overrightarrow{j_{m}})=(\mathsf{k}_{0},\overrightarrow{\mathsf{k}})=\mathsf{\mathsf{\mathsf{k}}}.\label{eq:59}\end{eqnarray}
Here we the symbol $(\leftarrow)$ stands the operation from right
to left and $\rho_{m}$ and $\overrightarrow{j_{m}}$are respectively
the charge and current source densities due to the dual charge given
by equations (\ref{eq:16}, \ref{eq:19} and \ref{eq:23} ). Equation
(\ref{eq:58}) may also be written as follows, \begin{eqnarray}
[\boldsymbol{\boxdot},\,\boldsymbol{\digamma_{\mu}}] & = & \mathbf{\boldsymbol{\mathsf{k}_{\mu}}}.\label{eq:60}\end{eqnarray}
Hence, we may write the analogous quaternion equation for the Lorentz
force equation of motion for dual charge given by equation (\ref{eq:29})
as

\begin{eqnarray}
g & [\boldsymbol{\mathsf{U}},\,\boldsymbol{\mathcal{\boldsymbol{F}}}] & =\boldsymbol{\mathbf{\Im}}\label{eq:61}\end{eqnarray}
where $\mathsf{U}$ and $\boldsymbol{\mathbf{\Im}}$ are respectively
the quaternionic analogues of four-velocity and four force given by
equations (\ref{eq:54}) and (\ref{eq:55}) and $[\boldsymbol{\mathsf{U}},\,\boldsymbol{\mathcal{\boldsymbol{F}}}]$
is the defined as the scalar product of quaternion velocity and quaternion
anti-symmetric field strength as

\begin{eqnarray}
[\boldsymbol{\boldsymbol{\mathsf{U}}},\,\boldsymbol{\mathcal{\boldsymbol{F}}}] & =\frac{1}{2} & (\overline{\boldsymbol{\boldsymbol{\mathsf{U}}}}\boldsymbol{\mathcal{\boldsymbol{F}}}+\overline{\boldsymbol{\mathcal{\boldsymbol{F}}}}\boldsymbol{\boldsymbol{\mathsf{U}}})\nonumber \\
= & U_{0}\boldsymbol{\digamma_{0}}+ & \overrightarrow{u}\cdot\boldsymbol{\overrightarrow{\digamma}}=(\mathsf{f}_{0},\overrightarrow{\mathsf{f}})=\mathsf{\mathsf{\mathsf{\boldsymbol{\mathbf{\Im}}}}}.\label{eq:62}\end{eqnarray}
Here $\mathsf{f}_{0}$ and $\overrightarrow{\mathsf{f}}$ are the
scalar and vector components of four-force given by equation (\ref{eq:29})
and thus we may write equation (\ref{eq:62}) like equation (\ref{eq:60})
as 

\begin{eqnarray}
[\boldsymbol{\mathbf{U}},\,\boldsymbol{\digamma_{\mu}}] & = & \mathbf{\boldsymbol{\mathsf{f}_{\mu}}}.\label{eq:63}\end{eqnarray}
Equations (\ref{eq:62}) and (\ref{eq:63}) may also be generalized
to the following quaternion form of Lorentz force equation of motion;

\begin{eqnarray}
\mathsf{\mathsf{\mathsf{\boldsymbol{\mathbf{\Im}}}}} & = & \boldsymbol{\mathsf{k}}\boldsymbol{\boxdot}\mathsf{B}\label{eq:64}\end{eqnarray}
where we have used the definition of the four-current density as $k_{\mu}=\rho_{m}U_{\mu}$(i.e
the product of charge density and velocity) and the volume integration
of charge density gives the total charge for point like monopole leading
to $g=\intop\rho_{m}d^{3}x$ resulting to establish the following
relation between the four current and four velocity as

\begin{eqnarray}
\mathbf{k}_{\mu} & = & g\,\mathbf{u}_{\mu}\label{eq:65}\end{eqnarray}
and the anti-symmetric field strength is replaced by the quaternionic
wave equation $\mathbf{\boldsymbol{\boxdot}}\mathbf{B}=\mathrm{\mathit{\mathsf{\mathfrak{\mathcal{\boldsymbol{\psi}}}}}}$
given by equation (\ref{eq:44}) to establish a connection between
potential and field in quaternion formulation. Equation (\ref{eq:64})
is same as derived earlier by Waser as the quaternionic form of generalized
Lorentz force and here we have obtained for the theories of dual electrodynamics.

\section{Discussion}

Starting with the concept of invariance of the electromagnetic duality
in source free Maxwell's equations (\ref{eq:1}) , we have shown that
the Maxwell's equations (\ref{eq:1}) remain invariant under duality
transformations (\ref{eq:2}, \ref{eq:3} ,\ref{eq:4}) with the choice
of four potential either to express the electromagnetic fields given
by equation (\ref{eq:8})or by (\ref{eq:10}). The conventional Maxwell's
equations (\ref{eq:7}) follow the first choice of four-potential
which describe the electromagnetic fields given by equation (\ref{eq:8})
and thus, violates the duality invariance due to presence of source
current. Here, we have discussed the alternative choice of second
four- potential which produces the electric and magnetic fields given
by equations (\ref{eq:10}) to satisfy the another kind of Maxwell's
equations given by equations (\ref{eq:11}) and (\ref{eq:12}). These
Maxwell's equations (\ref{eq:11}) or (\ref{eq:12}) are visualized
as dual Maxwell's equations as they may be obtained from the usual
Maxwell's equations (\ref{eq:7}) under duality transformations (\ref{eq:2},
\ref{eq:3} ,\ref{eq:4}) . So, it is concluded that the second alternative
potential is described as the dual electromagnetic potential and hence
produces the electric and magnetic fields for the dynamics of dual
electric charge ( i.e. magnetic monopole). As such, the duality transformations
(\ref{eq:13}) are established for four-potentials and four-currents
of electric and magnetic charges. It has been shown that electric
and magnetic charges and correspondingly the anti-symmetric electromagnetic
field tensor and its dual, transform under duality transformations
respectively given by equations (\ref{eq:14}) and (\ref{eq:15}).
Accordingly, we have discussed the Maxwell's equations (\ref{eq:19})
in terms of electromagnetic field vector and developed a covariant
formulation of parallel electrodynamics for dual electric charge (
magnetic monopole). Subsidiary conditions like Lorentz gauge condition
(\ref{eq:24}) and the continuity equation (\ref{eq:26}) are also
obtained consistently. We have also developed a connection between
the four potential and four currents given by field equation (\ref{eq:25})
and hence introduced a new vector parameter $\overrightarrow{S}$
as field current in the same manner to establish its relation with
the electromagnetic field vector $\overrightarrow{\psi}$ given by
equation (\ref{eq:27}). It has been concluded that like the dynamics
of electric charge, we may develop the parallel dynamics of dual charge
(magnetic monopole) and accordingly the equation of motion (\ref{eq:29})
has been established for Lorentz force equation of motion of magnetic
monopole. Thus, we observe that either the classical electrodynamics
(dynamics of electric charge -electron) or the dual electrodynamics
(dynamic of magnetic monopole) suffers from the fact that the classical
equations of motion are no more invariant under duality transformations.
So, in order to survive the duality invariance and to symmetrize the
Maxwell's equations, theories of dyons (particles carrying simultaneous
existence of electric and magnetic charges) do better and the bi-quaternion
formulation of generalized electromagnetic fields of dyons provide
a consistent platforms to understand the existence of monopole and
dyons. As such, we have discussed the bi-quaternionic formulation
of dual electrodynamics given by equations (\ref{eq:44} , \ref{eq:45},
\ref{eq:49}, \ref{eq:52}, \ref{eq:53}, \ref{eq:58}, \ref{eq:60},
\ref{eq:61} and \ref{eq:64}) and it has been shown that these quaternion
equations are compact, simple and manifestly covariant. It has been
shown that bi-quaternion reformulations of usual and dual electrodynamics
describe the change of metric from $(+,-,-,-)$ to $(-,+,+,+)$. Hence
it leads to the conclusion that on passing from usual Hilbert space
to a quaternionic space the signature of four - vector is changed
from $(+,-,-,-)$ to $(-,+,+,+)$. Hence the mapping $(3,1)\rightarrow(1,3)$
incorporated \cite{key-20} with complex superluminal Lorentz transformations
and the quaternionic formulation for subluminal field equations are
similar in nature. So, we may accordingly establish a closed connection
between bi-quaternion formulation and complex superluminal transformations.
The advantage in expressing the field equations in quaternionic forms
is that one may directly extend the theory of subluminal to superluminal
realm as well as the non commutativity of quaternion units play an
important role to understand the current grand unified theories and
the theories beyond the standard model of elementary particles. This
formalism may also be useful to develop space - time duality between
complex and quaternionic quantum mechanics such that the evolution
operator for bradyons depends on time and that for tachyons depends
on space. As such the bi-quaternion formulation may hope a better
understanding of duality invariance as the quaternion wave equations
represent their self dual nature. On the other hand, bi -quaternion
analyticity provides a unified and consistent grounds for the existence
of monopoles and dyons and here we have described the dual electrodynamics
accordingly. It may also be concluded that the quaternion formulation
be adopted in a better way to understand the explanation of the duality
conjecture and supersymmetric gauge theories as the candidate for
the existence of monopoles and dyons .

\begin{description}
\item [{Acknowledgment-}]~
\end{description}
The work is supported by Uttarakhand Council of Science and Technology,
Dehradun. One of us OPSN is thankful to Chinese Academy of Sciences
and Third world Academy of Sciences for awarding him CAS-TWAS visiting
scholar fellowship to pursue a research program in China. He is also
grateful to Professor Tianjun Li for his hospitality at Institute
of Theoretical Physics, Beijing, China.

\end{document}